\documentclass{iopart}

\usepackage{iopams}
\usepackage{graphicx}
\usepackage{color}
 
\newcommand{\irm}{{\rm i}}
\newcommand{\beq}{\begin{equation}}
\newcommand{\eeq}{\end{equation}}
\newcommand{\bdm}{\begin{displaymath}}
\newcommand{\edm}{\end{displaymath}}

\begin{document}

\title[]
{Suspension-thermal noise in spring-antispring systems for future gravitational-wave detectors}

\author{Jan Harms}
\vskip 1mm
\address{Universit\`a degli Studi di Urbino ``Carlo Bo'', I-61029 Urbino, Italy}
\author{Conor M Mow-Lowry}
\vskip 1mm
\address{Institute of Gravitational Wave Astronomy, School of Physics and Astronomy, University of Birmingham, Birmingham B15 2TT, UK
}

\begin{abstract}
Spring-antispring systems have been investigated as possible low-frequency seismic isolation in high-precision optical experiments. These systems provide the possibility to tune the fundamental resonance frequency to, in principle, arbitrarily low values, and at the same time maintain a compact design of the isolation system. It was argued though that thermal noise in spring-antispring systems would not be as small as one may naively expect from lowering the fundamental resonance frequency. In this paper, we present a detailed calculation of the suspension thermal noise for a specific spring-antispring system, namely the Roberts linkage. We find a concise expression of the suspension thermal noise spectrum, which assumes a form very similar to the well-known expression for a simple pendulum. It is found that while the Roberts linkage can provide strong seismic isolation due to a very low fundamental resonance frequency, its thermal noise is rather determined by the dimension of the system. We argue that this is true for all horizontal mechanical isolation systems with spring-antispring dynamics. This imposes strict requirements on mechanical spring-antispring systems for the seismic isolation in potential future low-frequency gravitational-wave detectors as we discuss for the four main concepts: atom-interferometric, superconducting, torsion-bars, and conventional laser interferometer.
\end{abstract}

%%%%% Body of the paper %%%%%%

\section{Introduction}
One of the important technologies enabling the detection of gravitational waves with ground-based laser interferometers is the seismic isolation of their test masses. Seismic isolation can generally be categorized into passive and active isolation. Active isolation utilizes seismic sensors to suppress seismic noise by means of feedback control or feedforward cancellation. A passive isolation is a contraption (typically purely mechanical) that does not contain any control loops to improve isolation performance. A pendulum is the simplest example of a passive seismic isolation with horizontal oscillations of the suspension point being suppressed at the suspended mass above the pendulum frequency. The Advanced LIGO seismic isolation system consists of several active and passive stages  \cite{MaEA2014,MaEA2015}, while the Advanced Virgo seismic isolation is mostly passive with active components targeting seismic noise only below the observation band \cite{BrEA2005,AcEA2010,AcEA2015}. 

Next-generation detectors are currently being planned in Europe, the Einstein Telescope \cite{PuEA2010}, and the US, the Cosmic Explorer \cite{AbEA2016d}. One of the goals of the Einstein Telescope is to extend the observation band to significantly lower frequencies, i.e., down to a few Hertz, which will require a new design of the seismic-isolation system. Better isolation performance at lower frequencies can in principle be achieved by increasing the dimension of the passive stages, and so the basic concept for the Einstein Telescope is to construct a larger version of the Virgo isolation system. Clearly, increasing the dimension of an isolation system has limits, and this strategy cannot be followed to provide sufficient seismic isolation below 1\,Hz for low-frequency GW detectors as discussed in \cite{HaEA2013}.

An alternative strategy is to make use of spring-antispring systems (SASs), where by design a partial cancellation of restoring forces is achieved to create a softer response, i.e., with lower fundamental resonance frequency \cite{Win2002}. Spring-antispring systems have been envisioned early-on as passive stages in vertical isolation chains where it is otherwise very challenging to realize soft supports of the load \cite{BrEA1993,BeEA1999}. Horizontal spring-antisprings were also installed as pre-isolation stage in Virgo in the form of inverted pendula \cite{LoEA1999}. A broader investigation of the isolation performance of SASs was carried out by Winterflood \cite{Win2002}, followed by detailed theoretical analyses of the Roberts linkage \cite{DuBl2010}. Proof-of-principle experimental demonstrations of the benefit of SASs have been carried out for the Scott-Russell linkage \cite{WLB1999}, which achieved a record-low fundamental resonance frequency of about 6\,mHz with $Q$-factor greater than 3, and Watt's linkage implemented in seismic sensors with the goal to improve sensitivity of the seismic measurement \cite{BaEA2016}. Spring-antispring dynamics can also be realized by combining mechanical and magnetic components. For example, the obsolete SS-1 Ranger seismometer formerly produced by Kinemetrics had small rod magnets positioned around the test mass suspended from a spring to increase the natural period by about a factor 3. The same principle is used in the Virgo Superattenuator to lower the fundamental resonance of the vertical stages \cite{BrEA2005}. Magnetically assisted suspension systems have also been proposed for millihertz GW detectors \cite{ThEA2017}.

With seismic noise being strongly suppressed, other forms of instrumental noise pose new sensitivity limitations. Among these, thermal noise of the suspension system is an important low-frequency contribution to instrumental noise. Since the suspensions are part of the seismic isolation system, suspension thermal noise and seismic isolation performance are not fully independent. The first calculations of thermal noise in suspension systems of gravitational-wave detectors were presented by Saulson \cite{Sau1990}. His analysis included systems that experience internal loss quantified by a loss angle, in addition to the widely applied viscous friction models. This is for example the case in well-designed, low-loss, fused-silica fibers that have been employed in the GEO600 and Advanced LIGO detectors, and will also be used in Advanced Virgo on their respective suspensions' lowest stage to reduce the impact of thermal noise. For fused-silica fibers, more careful analyses were performed incorporating the effects of gravity and elasticity in Lagrangian models \cite{GoSa1994,Gon2000}, and considering details of specific suspension systems \cite{HHR2004}. It was found that the effective mechanical loss in a simple fiber suspension can be described by introducing a damping dilution factor, which accounts for the combination of a lossless gravitational restoring force with a weakly anelastic restoring force from fiber bending \cite{RoEA1997,CaEA2000}.

These studies were accompanied by other publications focussing on specific aspects of loss mechanisms and their effect on the dynamics of a system. One analysis shows that the effect of mechanical loss on the dynamics of a mechanical system is enhanced if restoring forces partially cancel as is the case in SASs \cite{SaEA1994}. Specifically, the observed quality factor decreases with decreasing frequency of the fundamental resonance. It is therefore easier to observe anelastic properties of a material in SASs. Another important study focussed on the loss mechanism itself. Internal loss in suspension fibers can be due to material imperfections, or theromelastic damping due to irreversible heat flow across temperature gradients. A correct calculation of thermoelastic loss in fibers requires including the stress dependence of the loss angle, which can decrease (even cancel) or increase the overall thermoelastic loss \cite{CaWi2002}. 

In this paper, we present an analysis of thermal noise in spring-antispring systems; specifically, in the Roberts linkage. We introduce the formalism with two simple expamples. The simple pendulum is analyzed in section \ref{sec:simple}, and then the inverted pendulum in the form of a beam-balance tiltmeter as a simple spring-antispring system in section \ref{sec:tiltmeter}. The more complicated calculation for the Roberts linkage is shown in greater detail in section \ref{sec:roberts}. In section \ref{sec:apply}, we take the results to assess sensitivity limitations due to suspension-thermal noise in potential future ground-based, sub-Hz GW detectors, i.e., atom-interferometric GW detectors \cite{CaEA2016a}, superconducting GW detectors \cite{PaEA2016}, torsion-bar GW detectors \cite{ShEA2014,McEA2017}, and laser-interferometric GW detectors with conventional configuration \cite{HaEA2013}. All low-frequency concepts, except maybe for a novel atom-interferometric concept \cite{YuTi2011,GrEA2013}, require passive seismic isolation in the observation band as acknowledged in the respective references, and this cannot be achieved effectively without employing SAS components. We conclude in section \ref{sec:concl}.

\section{Suspension-thermal noise of a simple pendulum}
\label{sec:simple}
According to the fluctuation-dissipation theorem, the suspension thermal noise spectrum is determined by the real part of the complex admittance $Y(\omega)=-\irm\omega \tilde x(\omega)/\tilde F(\omega)$, where $\tilde x(\omega)$ is the displacement amplitude of a suspended test mass at frequency $\omega$ produced by a force $\tilde F(\omega)$ acting on it \cite{Kub1966,Sau1990}:
\beq
S_x(\omega)=\frac{4k_{\rm B}T}{\omega^2}\Re\left(Y(\omega)\right)
\label{eq:thermal}
\eeq
The admittance is calculated by solving the linearized equations of motion in frequency domain, and depends on the energy dissipation of the system under consideration. In mechanical suspension systems, the dissipation is dominated by the bending of weakly anelastic elements used in the suspension such as metallic wires or glass fibers.

For a simple pendulum of length $L$ and with suspended mass $m$, including the anelastic restoring force from fiber bending, the kinetic and potential energies $\mathcal T,\,\mathcal V$ in terms of the bending angle $\theta(t)$ read
\begin{eqnarray}
{\mathcal T} &= \frac{1}{2}m L^2\dot \theta(t)^2,\\
{\mathcal V} &= -mgL\cos(\theta(t))+\frac{1}{2}\tau L_{\rm el}\theta(t)^2+F(t) L\theta(t),
\end{eqnarray}
where $\tau=mg$ is the fiber tension, and $L_{\rm el}=\sqrt{E I_a/\tau}$ is the fiber bending length, i.e., the length from the fiber suspension point to the effective bending point \cite{CaEA2000}, which depends on the material's Young's modulus $E$, and on the fiber's second moment of area $I_a$ ($I_a=\pi r^4/2$ for cylindrical fibers of radius $r$). The elastic energy includes contributions from two bending points, i.e., the numerical factor is 1/4 for a single bending point. Note that this Lagrangian does not describe violin modes, i.e., the transverse vibration modes of the fibers. In this way, the only degree of freedom is given by the rotation angle $\theta(t)$. A Lagrangian that includes the violin modes was presented, for example, by Gonz{\'a}les and Saulson \cite{GoSa1994}.

Once translated into Fourier domain, the loss angle $\phi$ will be introduced through the Young's modulus via $E\rightarrow E(1+\irm\phi)$. So the relevant loss happens by producing strain in the fiber material. The potential energy already includes a contribution from the force $F(t)$ acting on the mass $m$, which is later used to calculate the admittance. Note that while the first term of the potential energy is valid for arbitrarily large angles $\theta$, the last two only yield accurate linearized equations of motion, and they take more complicated forms if non-linear effects are to be studied.

The equation of motion linearized with respect to the angle $\theta(t) $ then reads
\beq
F(t)/m+g(1+L_{\rm el}/L)\theta(t)+L\ddot\theta(t)=0
\eeq
It is now straight-forward to calculate the real value of the admittance by translating the equation into Fourier domain, introducing the loss angle, and using the substitution $\tilde\theta(\omega)\rightarrow\tilde x(\omega)/L$ valid for small displacements:
\beq
\Re(Y(\omega))=\frac{1}{m}\frac{\phi\omega\omega_{\rm el}^2/2}{\phi^2\omega_{\rm el}^4/4+(\omega_0^2+\omega_{\rm el}^2-\omega^2)^2},
\label{eq:admsimple}
\eeq
where $\omega_0^2=g/L$ and $\omega_{\rm el}^2=\omega_0^2L_{\rm el}/L$. The fraction $L_{\rm el}/L\ll 1$ is also known as damping dilution factor since it can formally be combined with the loss angle to obtain the effective loss $\phi_{\rm eff}=\phi L_{\rm el}/L$, in which case the last equation can be rewritten as
\beq
\Re(Y(\omega))=\frac{1}{m}\frac{\phi_{\rm eff}\omega\omega_0^2/2}{\phi_{\rm eff}^2\omega_0^4/4+(\omega_0^2-\omega^2)^2}.
\eeq
Here, we neglected the small change in resonance frequency due to the elastic restoring force. The suspension thermal noise is now determined by inserting the last equation into equation (\ref{eq:thermal}). Above resonance, the last equation can be approximated by
\beq
\Re(Y(\omega))\approx\frac{1}{m}\frac{\phi_{\rm eff}\omega_0^2/2}{\omega^3}.
\eeq
Since the transfer function of horizontal displacement of the suspension point to horizontal test-mass displacement, i.e., the seismic isolation, is $\mathcal H = -\omega_0^2/\omega^2$ above resonance, we find that seismic noise and suspension thermal noise can both be reduced by decreasing the pendulum resonance frequency $\omega_0$.

It can sometimes be helpful to consider thermal force fluctuations instead of position fluctuations. The thermal force spectrum assumes the form
\beq
S_F(\omega)=4k_{\rm B}T\Re\left(Z(\omega)\right),
\label{eq:thermalF}
\eeq
where $Z\equiv 1/Y$ is the impedance of the system. For a simple pendulum, the force spectrum is given by
\beq
S_F(\omega)=2k_{\rm B}Tm\phi\omega_{\rm el}^2/\omega.
\label{eq:pendF}
\eeq
This form shows that thermal fluctuations can be cast into a form such that they are connected to a pendulum's dissipation and only to its dissipation independent of other terms in the equation of motion. It also simplifies comparison with other forces acting on the test mass.

Finally, we can use the impedance to calculate the $Q$-factor of the pendulum using the method proposed in \cite{SaEA1994},
\beq
Q=\frac{\Re(-\irm\omega Z(\omega))\big|_{\omega=0}}{\Im(-\irm\omega Z(\omega))\big|_{\omega^2=\omega_0^2}^{\phantom{I}}}=\frac{2}{\phi}\frac{\omega_0^2+\omega_{\rm el}^2}{\omega_{\rm el}^{2^{\phantom{I}}}}=\frac{2}{\phi_{\rm eff}}.
\label{eq:Q}
\eeq
Here, the reason why the name ``damping dilution factor'' was given to the expression $\omega_0^2/\omega_{\rm el}^2$ becomes evident, and we want to mention here already that the Q value takes exactly the same form in the Roberts linkage with the respective definitions of $\omega_0$ and $\omega_{\rm el}$. We can express the complex admittance in terms of the $Q$-factor to obtain
\beq
\Re(Y(\omega))=\frac{1}{m}\frac{\omega\omega_0^2/Q}{(\omega_0^2/Q)^2+(\omega_0^2-\omega^2)^2},
\eeq
which has the advantage that parameters in this equation, i.e., the resonance frequency $\omega_0$, the $Q$-factor, and the mass $m$, can be measured relatively easily. This equation holds for all types of mechanical oscillators, but for more complicated systems, it might be necessary to calculate a model of the dynamics to determine the effective mass parameter, which does not always correspond to the total suspended load.

\section{Thermal noise of a beam-balance tiltmeter}
\label{sec:tiltmeter}
\begin{figure}[t]
\begin{center}
\includegraphics[width=0.95\textwidth]{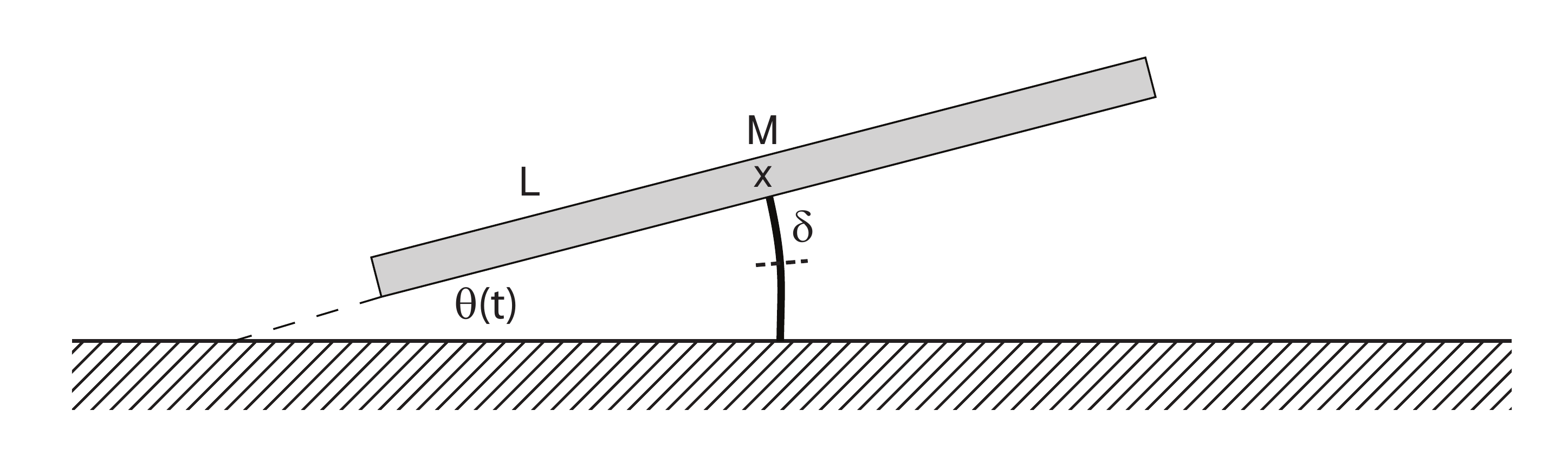}
\end{center}
\caption{Sketch of a beam-balance tiltmeter, with $\theta(t)$ being the rotation angle of the beam, $M$ being its mass, $L$ is half the length of the beam, and $\delta$ is the distance between the effective bending point of the flexure and the center of mass of the beam.}
\label{fig:tiltmeter}
\end{figure}

Beam-balance tiltmeters also show SAS dynamics, which are in fact equivalent to the dynamics of an inverted pendulum \cite{VeEA2014}. A beam of mass $M$ and moment of inertia $I_{\rm bm}$ is supported by a flexure. As shown in figure \ref{fig:tiltmeter}, we consider a one-dimensional model where the rotation of the beam is quantified by the angle $\theta(t)$. The center of mass (indicated by the cross) is located a distance $\delta$ above the effective bending point of the flexure. In the figure, the bending length of the flexure is slightly shorter than $\delta$ since the bending length is defined as the distance between the attachment point of the flexure to the beam and effective bending point. In practice though, the shape of the beam can be modified and the distance $\delta$ can in principle be made equal to the bending length. Note that the best beam-balance tiltmeters today have the beam supported from above so that the flexure is under tension instead of compression (see \cite{VeEA2014}), but this does not change the dynamics with respect to the simple model outlined in the following.

The kinetic and potential energies $\mathcal T,\,\mathcal V$ in terms of the rotation angle $\theta(t)$ can be cast into the form
\begin{eqnarray}
{\mathcal T} &= \frac{1}{2}I_{\rm bm} \dot\theta(t)^2,\\
{\mathcal V} &= Mg\delta\cos(\theta(t))+\frac{1}{2}\kappa\theta(t)^2+\zeta(t)\theta(t),
\end{eqnarray}
where $\kappa$ is bending stiffness of the flexure ($\kappa=EI_{\rm a}/L$ for a flexure of height $L$ and second moment of area $I_{\rm a}$, assuming that its stiffness is independent of the compressional force $Mg$), and $I_{\rm bm}$ the moment of inertia of the beam. It is tempting to introduce a bending length instead of the stiffness $\kappa$, but note that the elastic restoring force of the flexure is of different nature compared to the bending of a thin fiber, which means that one cannot immediately say what the equivalent of the fiber tension would be here.

In order to calculate the admittance, instead of using a force acting in normal direction to the beam on one of its ends, we apply a torque $\zeta(t)$ since the coordinate of interest is the rotation angle $\theta(t)$ and not so much the displacement of the end points of the beam.

Directly writing down the resulting real part of admittance,
\beq
\Re(Y(\omega))=\frac{1}{I_{\rm bm}}\frac{\phi\omega\omega_{\rm el}^2}{\phi^2\omega_{\rm el}^4+(\omega_0^2-\omega^2)^2},
\label{eq:admtilt}
\eeq
with $\omega_0^2=(\kappa-Mg\delta)/I_{\rm bm}$ and $\omega_{\rm el}^2=\kappa/I_{\rm bm}$, we obtain an expression that is formally similar to equation (\ref{eq:admsimple}). Substituting the admittance in equation (\ref{eq:thermal}) by this term, we get the angular thermal-noise spectral density $S_\theta$. The resonance frequency $\omega_0$ is determined by elastic as well as gravitational restoring forces, and it can be tuned to lower values by adjusting the center of mass.

Finally, the thermal torque-noise spectrum is given by
\beq
S_\zeta(\omega)=4k_{\rm B}TI_{\rm bm}\phi\omega_{\rm el}^2/\omega=4k_{\rm B}TI_{\rm bm}\omega_{\rm 0}^2/Q/\omega,
\eeq
where the $Q$-factor takes the same form as in equation (\ref{eq:Q}) without the factor 2 since the flexure stiffness is proportional to the Young's modulus, which means that we introduce the loss angle through $\kappa\rightarrow\kappa(1+\irm\phi)$.

\section{Suspension-thermal noise of a Roberts linkage}
\label{sec:roberts}
\begin{figure}[t]
\begin{center}
\includegraphics[width=0.95\textwidth]{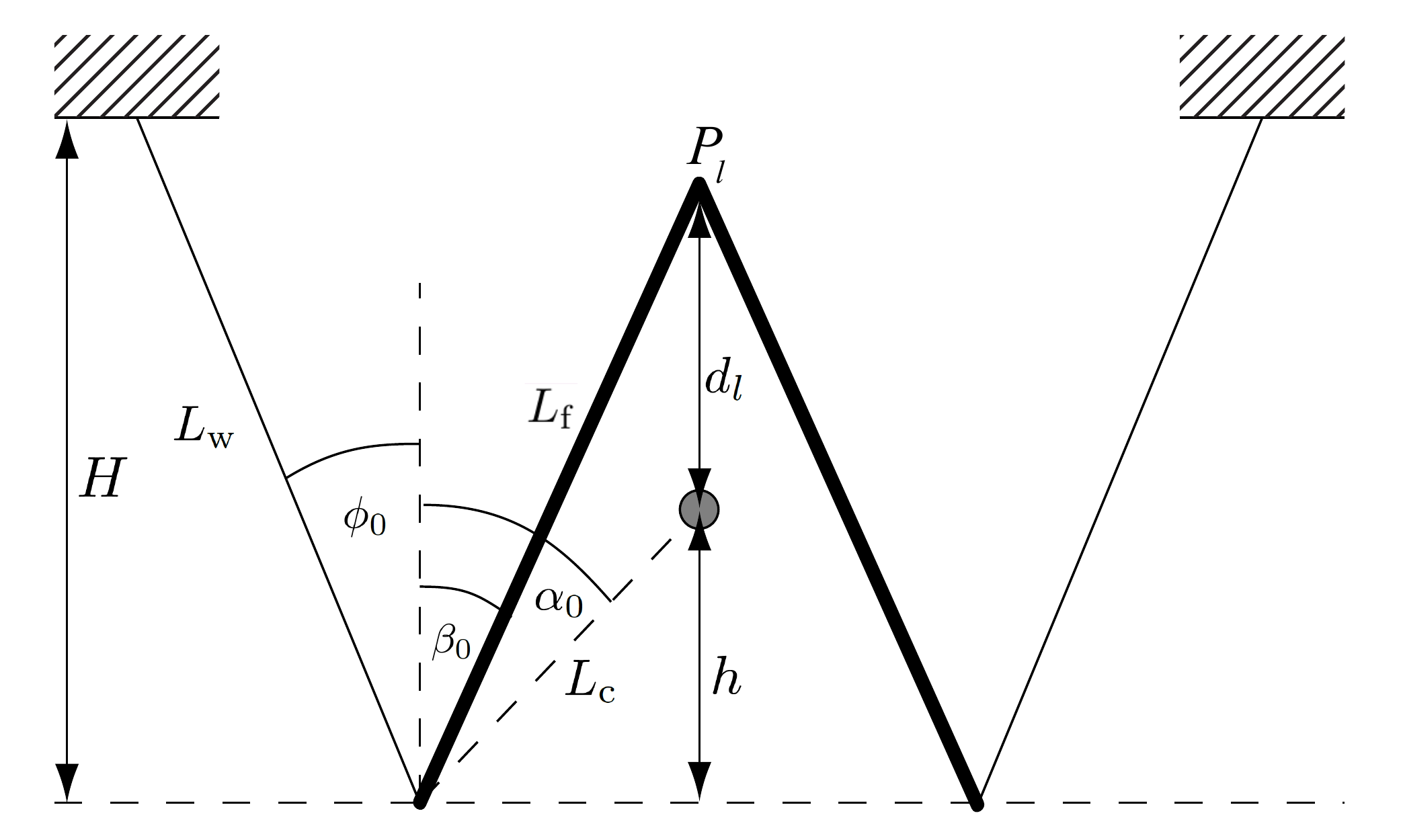}
\end{center}
\caption{Sketch of a Roberts linkage in equilibrium position. A rotation of the rigid frame relative to this position is parameterized by the angle $\theta(t)$.}
\label{fig:roberts}
\end{figure}

The calculation of the admittance of the Roberts linkage is analogous to the case of the simple pendulum. A simplified two-dimensional geometry of the linkage is sketched in figure \ref{fig:roberts}. A stiff frame represented by an isosceles triangle is suspended from flexible wires of equal length $L_{\rm w}$. 
The center of the mass $M$ of the frame is located a distance $d_{\rm l}$ below the vertex $P_{\rm l}$. The vertex corresponds to the suspension point of the load. In the following, we will assume that the load is a simple pendulum with suspended mass $m$. The pendulum is so short though that we can assume the load to be located vertically under the vertex at all times, i.e., the pendulum resonance frequency lies above the frequency band of interest. We will also neglect internal degrees of freedom of the linkage such as fiber violin modes and vibrations of the frame. This leaves one degree of freedom for the (two-dimensional) Roberts linkage, namely the rotation $\theta(t)$ of the frame.

In the following, we will largely adopt the notation of Dumas and Blair \cite{DuBl2010}. The positions of the center of mass of the frame and the load (effectively located at the vertex $P_{\rm l}$) can be written
\begin{eqnarray}
x_{\rm fr}(t) &= L_{\rm w}\sin(\phi_0+\theta(t))+L_{\rm c}\sin(\alpha_0+\theta(t)),\\
y_{\rm fr}(t) &= -L_{\rm w}\cos(\phi_0+\theta(t))+L_{\rm c}\cos(\alpha_0+\theta(t)),\\
x_{\rm l}(t) &= L_{\rm w}\sin(\phi_0+\theta(t))+L_{\rm f}\sin(\beta_0+\theta(t)),\label{eq:loadx}\\
y_{\rm l}(t) &= -L_{\rm w}\cos(\phi_0+\theta(t))+L_{\rm f}\cos(\beta_0+\theta(t)),
\end{eqnarray}
where $L_{\rm w}$ is the length of the wires supporting the frame, $L_{\rm c}$ is the distance between the lower suspension points of the frame to the frame's center of mass, and $L_{\rm f}$ is the length of the frame's ``leg''. The angle $\alpha_0$ is subtended by the arc between the frame's center of mass and the vertical from the frame's lower suspension point in equilibrium position, $\beta_0$ is the angle subtended by the arc between the frame's legs and the vertical, and $\phi_0$ is subtended by the arc between the frame's suspension wires and the vertical. The rotation angle of the frame from its equilibrium position is denoted by $\theta(t)$.

Deriving the equations of motion from the Lagrangian, we require the potential and kinetic energies
\begin{eqnarray}
{\mathcal T} &= \frac{1}{2}m (\dot x_{\rm l}(t)^2+\dot y_{\rm l}(t)^2)+\frac{1}{2}M (\dot x_{\rm fr}(t)^2+\dot y_{\rm fr}(t)^2)+\frac{1}{2}I_{\rm fr}\dot \theta(t)^2,\label{eq:kinetic}\\
{\mathcal V} &= F(t) x_{\rm l}(t)+mgy_{\rm l}(t)+Mgy_{\rm fr}(t)\nonumber\\
&\quad +\frac{1}{4}\left(\tau_{\rm w1} L_{\rm el,w1}+\tau_{\rm w2} L_{\rm el,w2}\right)\left((2\theta(t))^2+\theta(t)^2\right)+\frac{1}{4}\tau_{\rm l} L_{\rm el,l}\theta(t)^2,
\label{eq:potential}
\end{eqnarray}
where $\tau_{\rm w1}=((M+m)g/2)/\cos(\phi_0+\theta(t))$, $\tau_{\rm w2}=((M+m)g/2)/\cos(\phi_0-\theta(t))$, $\tau_{\rm l}=mg/\cos(\theta(t))$ are the tensions of the fibers supporting the frame and the load, $L_{\rm el,w1}=\sqrt{E I_a/\tau_{\rm w1}}$, $L_{\rm el,w2}=\sqrt{E I_a/\tau_{\rm w2}}$, $L_{\rm el,l}=\sqrt{E I_a/\tau_{\rm l}}$ are the resulting bending lengths assuming that all fibers have the same Young's modulus and area moment, and $I_{\rm fr}$ is the moment of inertia of the frame. Note that the bending angles at the attachment points to the frame are $2\theta(t)$ since the frame counter-rotates with respect to the suspension fibers by the same angle. It is then straight-forward to calculate the admittance of the Roberts linkage
\beq
\Re(Y(\omega))=\frac{(d_l+h+H)^2}{I_0}\frac{(\phi/2)\omega\omega_{\rm el}^2}{(\phi/2)^2\omega_{\rm el}^4+(\omega_0^2+\omega_{\rm el}^2-\omega^2)^2},
\label{eq:admitroberts}
\eeq
defining
\begin{eqnarray}
I_0&=I_{\rm fr}+(d_l+h+H)^2m+(h+H)^2M,\\
\omega_0^2&=\frac{g((H-h)(m+M)-d_lm)}{I_0},\label{eq:robertsres}\\
\omega_{\rm el}^2&=\frac{1}{2I_0}\left(\tau_lL_{\rm el,l}+10\tau_wL_{\rm el,w}\right)\Big|_{\theta=0}\label{eq:omegael}.
\end{eqnarray}
According to these results, the admittance can be written in a form very similar to the one of a simple pendulum; see equation (\ref{eq:admsimple}). Formally, the main difference is that the mass $m$ of the simple pendulum is substituted by $I_0/(d_l+h+H)^2$. 

We also present the result for the thermal force fluctuations acting on the suspended mass of a Roberts linkage. Evaluating equation (\ref{eq:thermalF}), one obtains
\beq
S_F(\omega)=2k_{\rm B}T\frac{I_0}{(d_l+h+H)^2}\phi\omega_{\rm el}^2/\omega,
\eeq
which is formally very similar to equation (\ref{eq:pendF}), with the suspended mass $m$ of the simple pendulum substituted by the effective mass $I_0/(d_l+h+H)^2$. 

Using the impedance to calculate the $Q$-factor of the Roberts linkage, we obtain exactly the same expression as in equation (\ref{eq:Q}), which means that the damping dilution factor also plays a role in SASs. However, as we have already seen, it does not appear in the expression for the suspension-thermal noise of Roberts linkages. Generally, for thermal noise, it is not the $Q$-factor, but geometric relations connecting test-mass displacement and bending angles that are relevant.

So as to minimize high-frequency horizontal translation of the load due to horizontal vibration of the linkage's suspension points, one must tune the center of percussion \cite{JuBl1994} so that it coincides with the suspension point of the load by setting $d_l=I_{\rm fr}/(h+H)M$ \cite{DuBl2010}. In this way, horizontal high-frequency vibration of the linkage's suspension points transfers predominantly into frame tilt around the load suspension point. Even when applying this parameter constraint, it is possible to tune the system parameters such that the resonance frequency $\omega_0$ approaches 0 until practical limitations are met or ultimately nonlinear dynamics become significant. Note that in contrast to the simple pendulum, it is not possible to express the elastic resonance frequency $\omega_{\rm el}$ in terms of the pendulum frequency $\omega_0$, since bending angles are determined by the system's dimension, while the resonance frequency is not fully determined by the dimension, but rather by the fine-tuning of certain parameters. This has important implications for the thermal noise as we are going to explain next.

Since the resonance frequency $\omega_0$ needs to lie well below the observation band so that the suspension system provides seismic isolation, one is mostly interested in thermal noise well above $\omega_0$, where admittance can be approximated by
\beq
\Re(Y(\omega))\approx\frac{(d_l+h+H)^2}{I_0}\frac{\phi\omega_{\rm el}^2/2}{\omega^3}
\label{eq:admhigh}
\eeq
This expression scales with $1/L^2$, where $L$ is a characteristic length scale of the system. Therefore, while seismic isolation performance improves when tuning the SAS to achieve lower $\omega_0$, the last equation shows that thermal noise is unaffected by the tuning, and stays at a level characteristic for a system of size $L$. It is clear then that the simple relation between $\omega_0$ and $\omega_{\rm el}$ for the simple pendulum is not generically true for all isolation systems, and SASs suffer from comparatively high thermal fluctuations of the test-mass position due to their compactness. 

\begin{figure}[t]
\begin{center}
\includegraphics[width=0.9\textwidth]{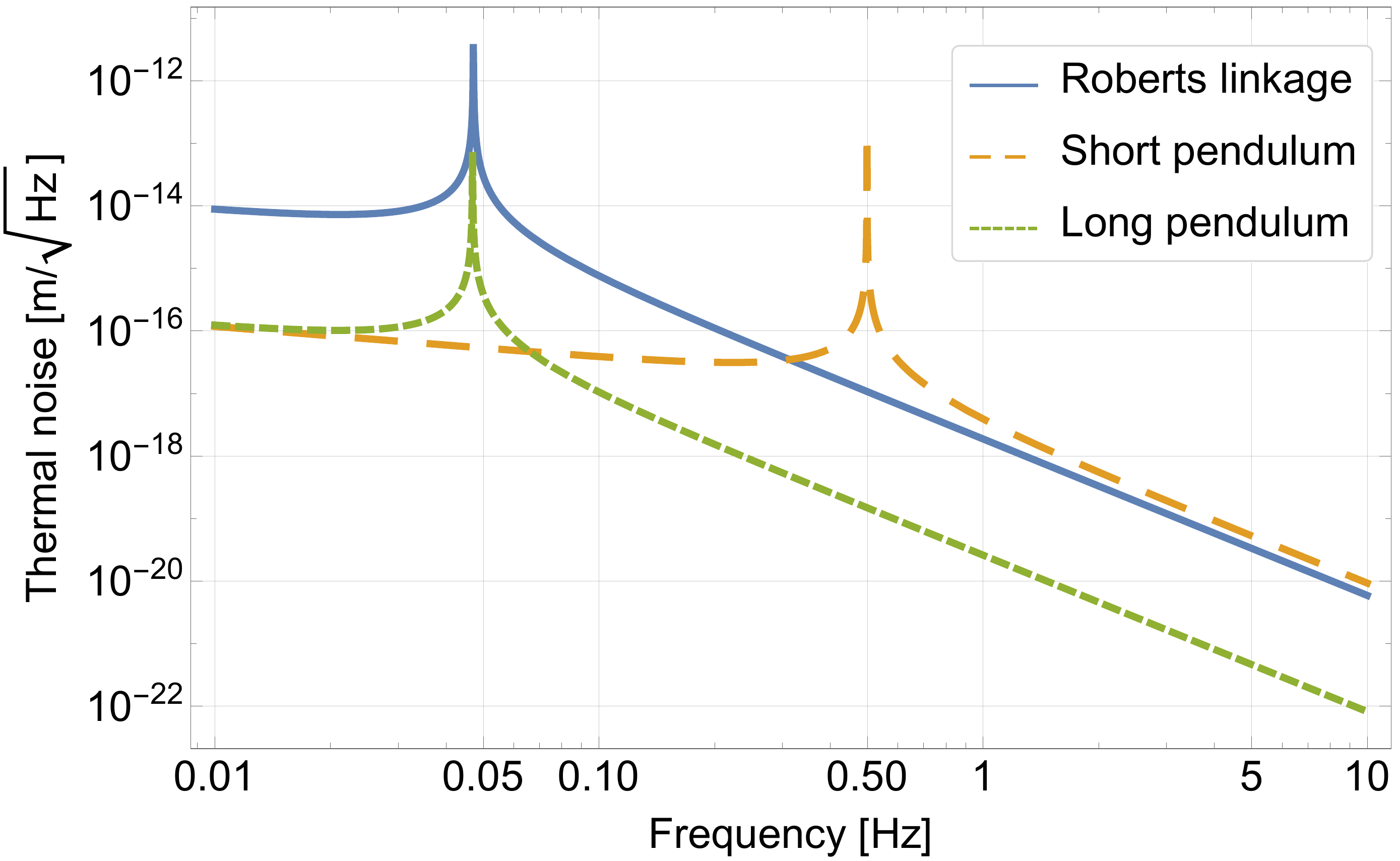}
\end{center}
\caption{Comparison of thermal-noise spectra of a Roberts linkage and two simple pendula in units of square-root of power spectral density. The long pendulum has the same resonance frequency, while the short pendulum has the same dimension of the Roberts linkage. Thermal noise is calculated with respect to the position fluctuations of a suspended test mass.}
\label{fig:thermalnoise}
\end{figure}
We can illustrate this further by plotting the corresponding thermal-noise spectra as shown in figure \ref{fig:thermalnoise}. The parameters used in this plot are summarized in table \ref{tab:thermal}. The loss angle is an estimate obtained by extrapolating models fit to higher-frequency data to sub-Hz frequencies \cite{HeEA2014}.
\begin{table}[ht!]
\renewcommand\arraystretch{1.15}
\begin{center}
\begin{tabular}{|c|c||c|c||c|c|}
\hline
\multicolumn{2}{|c||}{\bf Roberts linkage} & \multicolumn{2}{|c||}{\bf Long pendulum} & \multicolumn{2}{|c|}{\bf Short pendulum}\\
\hline
Parameter & Value & Parameter & Value & Parameter & Value \\
\hline
$H$ & 1\,m & $L$ & 113\,m & $L$ & 1\,m \\
\hline
$h$ & 0.85\,m &&&&\\
\hline
$d_{\rm l}$ & 0.135\,m &&&&\\
\hline
$\cos(\phi_0)$ & 0.98 &&&&\\
\hline
$M$ & 100\,kg &&&&\\
\hline
$I_{\rm fr}$ & 25\,kg\,m$^2$ &&&&\\
\hline
\multicolumn{6}{|c|}{\bf Common parameters}\\
\hline
\multicolumn{3}{|c|}{$m$} & \multicolumn{3}{|c|}{300\,kg} \\
\hline
\multicolumn{3}{|c|}{$E$ (fused silica)} & \multicolumn{3}{|c|}{72\,GPa} \\
\hline
\multicolumn{3}{|c|}{$r$ (fiber radius)} & \multicolumn{3}{|c|}{0.5\,mm} \\
\hline
\multicolumn{3}{|c|}{$\phi$} & \multicolumn{3}{|c|}{$10^{-7}$} \\
\hline
\multicolumn{3}{|c|}{$T$} & \multicolumn{3}{|c|}{300\,K} \\
\hline
\end{tabular}
\caption{Parameters used to calculate the thermal-noise spectra in figure \ref{fig:thermalnoise}.}
\label{tab:thermal}
\end{center}
\end{table}
We compare the thermal noise of a Roberts linkage with the noise of a long pendulum that has the same pendulum resonance frequency, and with a short pendulum of length $H$ corresponding to the height of the Roberts linkage. Above resonance, the thermal noise of the linkage is similar to the thermal noise of the short pendulum. One can understand this also intuitively since the mechanical loss is determined by how much the fibers are bent. The bending angles are characteristic of a system's dimension, not its fundamental resonance frequency. 

Equation (\ref{eq:admhigh}) shows several important scaling relations for suspension-thermal noise power spectral density above resonance. There is a linear dependence on the temperature $T$, and on the loss angle $\phi$. For the dimension, $L$, of the system, we find the typical $1/L^2$ scaling. The mass dependence is slightly more complex. Equation (\ref{eq:admhigh}) contributes a factor inversely proportional to mass through the moment of inertia, $I_0$. However, $\omega_{\rm el}$ also depends on the mass. According to equation (\ref{eq:omegael}), the fiber tension and moment of inertia have canceling mass dependence. To calculate the mass scaling of the bending length, we assume that the fiber cross section will always be minimized to support a certain load. The bending length then depends on the maximal stress $\sigma_{\rm yield}$ that the fiber can support, and it can be written as $L_{\rm el}\sim r_{\rm min}\sqrt{E/\sigma_{\rm yield}}$, where $r_{\rm min}$ is the minimal radius of the fiber, which scales as $m^{1/2}$. Bringing all mass terms together, we find that suspension-thermal noise scales as $1/m^{1/2}$. This analysis also shows that strong (high $\sigma_{\rm yield}$) and soft (low $E$) materials produce the smallest thermal noise. We point out though that there are other ``hidden'' mass dependencies. For example, suspending a very heavy load might limit the choice of materials that can be used for the suspension, which can affect thermal noise. Also, the loss angle depends on mass if the dominant dissipation mechanism is thermoelastic damping \cite{CaWi2002}.
\begin{table}[ht!]
\renewcommand\arraystretch{1.5}
\begin{center}
\begin{tabular}{|c|p{5cm}|}
\hline
Suspended mass & $\propto 1/m^{1/2}$ (neglecting mass-dependent loss angle and constraints on material choice) \\
\hline
Dimension & $\propto 1/L^2$ \\
\hline
Dissipation & $\propto \phi$ \\
\hline
Temperature & $\propto T$ \\
\hline
Strength & $\propto 1/\sigma_{\rm yield}^{1/2}$ \\
\hline
Stiffness & $\propto E^{1/2}$ \\
\hline
\end{tabular}
\caption{Parameter-scaling relations of suspension-thermal noise spectral densities valid for simple pendula and Roberts linkages.}
\label{tab:scalings}
\end{center}
\end{table}

The horizontal seismic isolation provided by a Roberts linkage is described by the horizontal response of the suspended load to seismic vibrations at the fiber suspension points. Ignoring small elastic restoring forces, the horizontal response of a test-mass suspended from a Roberts linkage to horizontal seismic displacement common between the two suspension points is given by
\beq
{\mathcal H}=\frac{\omega_0^2-(I_{\rm fr}-d_{\rm l}(h+H)M)\omega^2/I_0}{\omega_0^2-\omega^2}
\label{eq:isolroberts}
\eeq
The second term in the numerator is a result of the center-of-percussion effect, which is relevant in any suspension system where loads need to be considered as extended objects. In Roberts linkages, it sets a limit to how much seismic noise can be suppressed at high frequencies. One can choose masses and geometry to cancel this term, but there are practical limitations to how accurately the parameter values can be designed, and even a change of geometry due to temperature changes may be significant.
\begin{figure}[t]
\begin{center}
\includegraphics[width=0.9\textwidth]{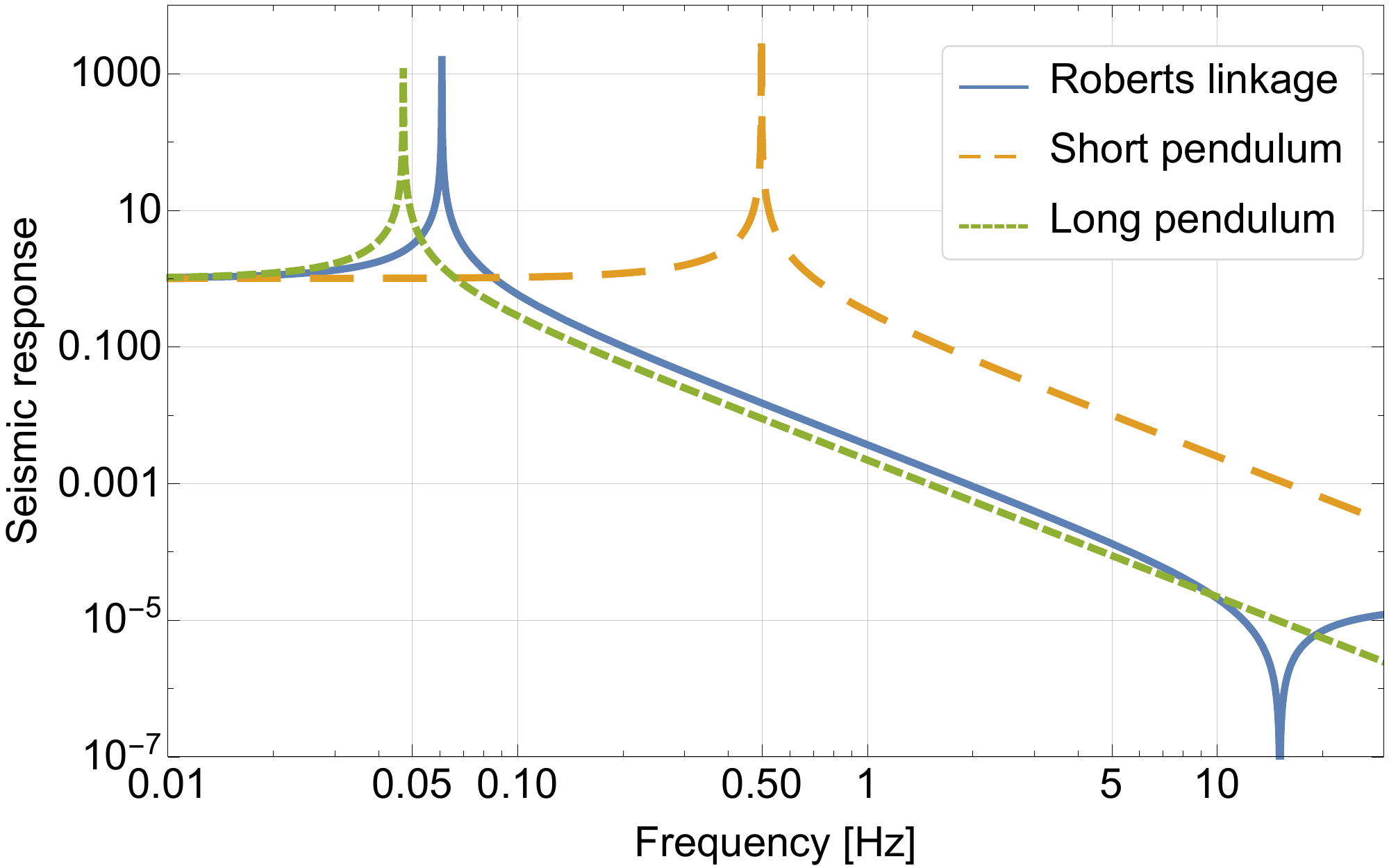}
\end{center}
\caption{Comparison of seismic-isolation performances. The long pendulum has a similar resonance frequency, while the short pendulum has the same dimension of the Roberts linkage.}
\label{fig:isolation}
\end{figure}

The isolation function in equation (\ref{eq:isolroberts}) is plotted in figure \ref{fig:isolation} using the parameter values shown in table \ref{tab:thermal}, elastic forces included, in comparison with the isolation functions of two simple pendula. It shows that a compact Roberts linkage can in principle provide the seismic isolation of a pendulum with more than 100\,m length, which is why SASs are interesting for applications in low-frequency GW detectors.

Let us briefly discuss the application of our results to other SASs.  Provided that the dominant dissipation process is bending of weakly anelastic fibers, all parameter scalings will be preserved. If the dominant dissipation process is different, then the definition of $\omega_{\rm el}$ in equation (\ref{eq:omegael}) needs to be modified, which can change parameter scalings. Assuming similar values of the fundamental resonance frequency, suspended mass, dimension, temperature, and material parameters of the fibers, different mechanical SAS systems have similar thermal-noise spectra. To be more careful, we also need to assume that the suspension system is well-designed, which means that unwanted cross-couplings between degrees of freedom are small, stresses in the bending fibers are solely to counteract the suspended weight, and bending angles are all similar and can be estimated from the total dimension of the SAS. Well designed pendula and SASs such as Scott-Russell linkages, or Watt linkages used as horizontal seismic isolation fulfill these requirements. 

Vertical seismic-isolation systems can be treated with the same formalism, but the bending components are typically flexures and cantilevers. We have already seen in the example of the tiltmeter that elastic restoring forces have a different nature in flexures and cantilevers than in thin suspension fibers. This changes some of the parameter scalings, but thermal-noise spectra can be calculated using the same formulas, as for example equation (\ref{eq:admitroberts}). We refer to Winterflood \cite{Win2002} for a more detailed discussion of SAS vertical seismic-isolation systems.

\section{Implications for low-frequency seismic isolation systems}
\label{sec:apply}
The goal of the following analysis is to determine whether suspension-thermal noise might pose a sensitivity limitation in sub-Hz GW detector concepts. We therefore assume that seismic noise is suppressed to a level below the suspension-thermal noise, which is of course a major challenge. The envisioned observation band of these detector concepts lies between about a few tens of mHz and a few Hz \cite{HaEA2013}. It should be noted that a way to reduce suspension-thermal noise relative to the level shown in figure \ref{fig:thermalnoise} is to implement cryogenic technology to lower the temperature of the suspensions. For this reason, the estimates of how suspension-thermal noise will limit the sensitivity of low-frequency detectors provided in this section might be considered conservative. Table \ref{tab:numbers} lists a few parameter values relevant to the following discussion.
\begin{table}[ht!]
\renewcommand\arraystretch{1.3}
\begin{center}
\begin{tabular}{|p{7.5cm}|c|}
\hline
Parameter & Value at 0.1\,Hz\\
\hline
Typical ground motion at quiet site & $5\cdot 10^{-7}\,\rm m/\sqrt{Hz}$ \\
\hline
Noise of best commercial seismometer & $5\cdot 10^{-10}\,\rm m/\sqrt{Hz}$ \\
\hline
Suspension-thermal noise at $T=300$\,K and for an effective suspended mass of a few 100\,kg & $10^{-15}\,\rm m/\sqrt{Hz}$ \\
\hline
Typical GW sensitivity target & $10^{-20}\,\rm Hz^{-1/2}$ \\
\hline
Seismic-noise suppression by a 1\,mHz passive filter  & $10^4$ \\
\hline
Predicted common-mode rejection in Bragg-type atom-interferometric gravity gradiometers with several 100\,m baseline & $\sim 10^4$ \\
\hline
Best past common-mode rejection in 10\,cm-scale superconducting gravity gradiometers  & $3\cdot 10^7$ \\
\hline
Horizontal displacement to differential rotation cross-coupling expected in current torsion-bar gravity gradiometers  & $\lesssim 0.01$ \\
\hline
Typical cross-coupling in large-scale GW detectors (gravity induced) & $\lesssim 0.001$ \\
\hline
\end{tabular}
\caption{Parameter values relevant to seismic isolation and suspension-thermal noise.}
\label{tab:numbers}
\end{center}
\end{table}

\paragraph{Seismic isolation} Generally, seismic isolation needs to be provided for more than one degree of freedom, since there will always be some level of cross-coupling between them \cite{AcEA2010,MaEA2014,MaEA2015}. This means that the resonance frequencies of a few relevant degrees of freedom in the seismic response of the isolation system must all lie below the observation band, and it is far from obvious that it can be achieved in the foreseeable future for low-frequency detectors \cite{Win2002}. 

The analysis in this paper will be simplified by treating the seismic isolation as a one-dimensional system. Proper modeling of cross-couplings between degrees of freedom is challenging since they often depend on accuracy limitations of the production process of a suspension system, which can only be estimated. 

Assuming that seismic noise can be suppressed sufficiently, suspension-thermal noise might become the new limiting noise source. With typical ground displacement of order $10^{-7}\,\rm m/\sqrt{Hz}$ at 0.1\,Hz, this can only be achieved with a multi-stage seismic isolation, possibly starting with an active stage that would ideally reduce the noise to a few times $10^{-10}\,\rm m/\sqrt{Hz}$ corresponding to the instrumental noise at 0.1\,Hz of the most sensitive broadband seismometers today, followed by passive SAS stages. While all stages have in principle an effect on the suspension-thermal noise, one can obtain a fairly accurate estimate by considering only the ultimate passive stage. This is only true though if the suspension system is designed appropriately. In general, and for a careful analysis, propagation of thermal noise from upper stages and coupling between stages need to be taken into account. Nonetheless, in the following, we will only consider the final suspension stage and use the solid curve in figure \ref{fig:thermalnoise} as thermal-noise spectrum.

\paragraph{Laser-interferometric GW detectors with conventional configuration} By conventional configuration, we mean a laser interferometer similar to the existing large-scale GW detectors. It turns out that this low-frequency concept is most strongly affected by suspension-thermal noise as already anticipated by Harms et al \cite{HaEA2013}. This is because unlike the other three concepts, there is generally no common-mode rejection mechanism concerning test-mass displacement noise \footnote{There will be common-mode rejection of seismic noise at upper stages, but these are not effective relative to suspension-thermal noise at the ultimate suspension stage.}. Assuming a mechanical system, the ultimate suspension stage needs to be a SAS stage since otherwise, e.g., with a short simple pendulum, the GW response would be strongly suppressed. The estimation of suspension-thermal noise in units of differential gravity strain is therefore simple: multiply the displacement noise $10^{-15}\rm\,m/\sqrt{Hz}$ at 0.1\,Hz, as predicted by our analysis with parameter values listed in table \ref{tab:numbers}, by $2/L$, where $L$ is the separation of test masses in one interferometer arm. 

If arm lengths of order 1\,km are assumed, then suspension-thermal noise would have to be reduced by another 2 orders of magnitude at least. It is clear that according to the scaling relations in table \ref{tab:scalings}, gaining 2 orders of magnitude even with a combination of modifications is extremely challenging (note that this translates into 4 orders of magnitude with respect to the noise spectral density, which is the reference of the scaling relations), and we consider it unfeasible in the foreseeable future.

\paragraph{Atom-interferometric GW detectors} The basic scheme of atom-interferometric GW detectors is to realize a differential laser-phase measurement over a long baseline using freely-falling, ultra-cold atoms. Interactions of the atoms with laser beams force each atom to move in a superposition of two paths forming the atom interferometer. Atoms are then counted in the output ports of the atom interferometer, and the atom number in each output is sensitive to the laser phase. 

The fact that the test masses consist of freely-falling atoms provides some immunity to seismic noise, but any form of displacement noise of the laser optics couples into the atom-interferometric readout. If the atom interferometer is interrogated in the Bragg regime with interrogating laser beams propagating in Fabry-Perot cavities, then optics displacement noise $\delta x(\omega)$ adds to the GW strain signal as \cite{CaEA2017}
\beq
h_{\delta x}(\omega) = \frac{2\sqrt{2}\,\omega}{\omega+\omega_{\rm p}}\frac{\delta x(\omega)}{L},
\eeq
with the cavity pole frequency $\omega_{\rm p}=\pi c/(2L\mathcal{F})$. The pole frequency lies well above the observation band of atom-interferometric GW detectors, in which case the last equation simplifies to 
\beq
h_{\delta x}(\omega) \approx 4\sqrt{2}\frac{\omega \,\delta x(\omega)}{c}\frac{\mathcal{F}}{\pi}.
\eeq
With a cavity finesse of a few hundred, we obtain
\beq
h_{\delta x}(0.1\,{\rm Hz}) \approx 10^{-6}\,{\rm m}^{-1}\delta x(0.1\,\rm Hz).
\eeq
If displacement noise of the cavity optics is dominated by thermal-suspension noise, then using the parameter values of the previous examples, we have $\delta x(0.1\,\rm Hz)\approx 10^{-15}\,\rm m/\sqrt{Hz}$ (see figure \ref{fig:thermalnoise}), which leads to a strain noise of $h_{\delta x}(0.1\,{\rm Hz})\approx 10^{-21}\rm Hz^{-1/2}$. With sensitivity goals at about $10^{-20}\rm Hz^{-1/2}$, suspension-thermal noise would not be a major noise contribution. However, this will depend on the ultimate choice for the cavity finesse, and also whether loss angles as small as $10^{-7}$ can be realized in a Roberts linkage (or similar SASs).

\paragraph{Superconducting GW detectors} Superconducting GW detectors measure differential displacements of levitated test masses against a common, rigid, reference frame, which must be suspended for the purpose of seismic isolation. All noise originating from the suspension system, e.g., thermal noise, seismic noise, or crackle noise \cite{Lev2012}, is suppressed by common-mode rejection in the differential readout of test-mass displacements \cite{MPC2002}. 

The rejection is first of all limited by the alignment accuracy of readout axes, adjustments in the readout system, and sensitivity and linear dynamic range of the test-mass displacement sensors \cite{MCP1986}. Ultimately, the maximally achievable common-mode rejection depends on the stiffness, and potentially symmetry of the reference frame and therefore becomes an issue especially in designs with larger frames. 

Stable common-mode rejections of $3\cdot 10^7$ are achieved today in three-axis, diagonal-component superconducting gravity gradiometers \cite{PaEA2016}. This means that a $10^{-15}\rm\,m/\sqrt{Hz}$ displacement noise at 0.1\,Hz would be suppressed to a value of less than $10^{-22}\rm\,m/\sqrt{Hz}$, which is sufficient for sub-Hz GW detection. It therefore seems that suspension-thermal noise will not be an issue in superconducting GW detectors. 

We still consider this target ambitious, since rejection ratios of order $10^8$ have only been achieved in smaller-scale experiments so far, i.e., several tens of centimeters, and it is possible that such high rejection ratios cannot be maintained when frame dimensions are increased for a full-scale GW detector. It has to be kept in mind as well that the reference frame plus test masses of a superconducting GW detector are proposed with a mass of several thousand kilograms \cite{PaEA2016}. In this case, it is possible that the suspensions have a loss angle significantly higher than $10^{-7}$ if for example metal wires instead of fused-silica fibers need to be used. Also, especially in such high-mass systems, recoil loss due to coupling between the suspended frame and ``upper stages'' of the suspension system, which in principle even includes the ground supporting the experiment, might be significant with possible degradation of mechanical quality factors.

\paragraph{Torsion-bar GW detectors} Torsion-bar antennas observe GWs by measuring the differential rotation of two orthogonal bars suspended from (torsion) fibers. Ideally, seismic or thermal noise entering through the suspension points of the torsion fibers leads to common rotations or displacements of the bars and is therefore rejected. So the main coupling mechanism is produced by (unavoidable) inaccuracies  of the suspension and bar design leading to asymmetries in the system and coupling between degrees of freedom. Current state-of-the-art is to add movable masses to the bars to be able to balance the system and tune various resonance frequencies \cite{McEA2016}. 

According to a model presented in McManus et al \cite{McEA2017}, cross coupling between horizontal motion of the torsion-fiber suspension point and differential bar rotation is currently estimated to be at the level $\lesssim 0.01$. Cross-coupling in current large-scale detectors is at a level $\lesssim 0.001$, but this coupling is dominated by the mismatch of directions of local gravity at different test masses of order $L/R_\oplus$, where $L$ is the arm length of the detector, and $R_\oplus$ the radius of Earth, and inaccuracies in the mechanical design might well contribute significantly lower cross-coupling. In torsion-bar detectors, a rejection by a factor greater than $10^4$ is required to suppress suspension-thermal noise according to figure \ref{fig:thermalnoise} below  $10^{-20}\rm\,Hz^{-1/2}$ at 0.1\,Hz (assuming 10\,m bars in future experiments). This seems extremely challenging and certainly requires more sophisticated balancing methods than the mechanical tuning used today.

It should be noted that thermal noise from a vertical seismic isolation would be less relevant since the same cross-coupling model predicts that vertical motion of the suspension point of the torsion fibers couples into differential bar rotation at a level $\sim 10^{-8}$. This should be sufficient even when taking into account that vertical isolation is expected to have a much larger loss angle, e.g., characteristic for metallic cantilevers. 

\section{Conclusion}
\label{sec:concl}
In this paper, we have presented a calculation of suspension-thermal noise of a Roberts linkage, which is a spring-antispring system (SAS) that might be used in future GW detectors to provide horizontal seismic isolation. We then analyzed how suspension-thermal noise contributes in various ground-based, sub-Hz GW detector concepts, and found that its effect ranges from probably insignificant noise to a potential show stopper. 

It was shown that an SAS by itself, even when engineered with very low fundamental resonance frequency, does not help to lower suspension-thermal noise. Thermal-noise spectra above the fundamental resonance are determined by the dimension of the SAS, with higher thermal noise for a smaller dimension, and does not directly depend on the fundamental resonance frequency. It can be understood intuitively by realizing that the mechanical damping responsible for the thermal noise occurs with weakly anelastic deformations of suspension fibers, which are related to bending angles, and therefore to a system's size. This revises common understanding that damping dilution, which is a factor in the relation between loss angle and $Q$-factor, generally plays a role in thermal noise.

Therefore, if severe sensitivity limitations by suspension-thermal noise are expected, then the only way to solve the problem is (1) to lower the temperature of the suspensions, (2) to lower the mechanical loss, (3) to decrease the elastic bending length by using lower Young's modulus materials and/or higher stress, or (4) to increase the dimension of the isolation system.

While the analysis focused on the Roberts linkage, qualitatively, we argued that the results hold for all (well-designed) mechanical SASs used for horizontal seismic isolation assuming dominant dissipation through the bending of thin fibers, which includes the Watt linkage and the Scott-Russell linkage. We have not investigated the case of vertical seismic isolation in detail. Since the dissipation in mechanical systems typically occurs through bending of cantilevers, which is a different type of elastic restoring force compared to the bending of thin fibers, the scaling of thermal-noise spectra for example with mass can change, but formally the calculation remains the same. It would be interesting to investigate how stress in flexures and cantilevers modify the stiffness of bending components (which is well understood for fibers), so that accurate thermal noise spectra can be calculated for vertical isolation systems.

%%%%%%%%%%%%%%%%%%%%%%%%%%%%%%%%
\ack{We thank Matthew Evans for discussions about spring-antisprings, which led to the writing of this paper. We also want to thank Yuri Levin and Chris Wipf who helped to improve our understanding of certain aspects of this work. We are also grateful to Jeff Kissel for detailed comments on the manuscript. This project has received funding from the European Union's Horizon 2020 research and innovation programme under the Marie Sklodowska-Curie grant agreement No 701264.}

\section*{References}
%\bibliographystyle{iopart-num}
%\bibliography{C:/MyStuff/references}
\providecommand{\newblock}{}

\end{document}